\PassOptionsToPackage{pdfpagelabels=false}{hyperref} 
\documentclass[usenatbib,fleqn]{mnras}
\usepackage{fixltx2e,longtable,ifpdf,amsmath,amssymb}
\usepackage{times,nicefrac,bm}
\ifpdf
  \usepackage{aeguill}
  \usepackage[pdftex]{graphicx,color}
\else
  \usepackage[dvips]{graphicx,color}
\fi

\newcommand{\teff}{\ensuremath{T_{\rm eff}}}
\newcommand{\logg}{\ensuremath{\log{g}}}

\newcommand{\kms} {\ensuremath{\mbox{km}\;\mbox{s}^{-1}}}

\def\utw{\smash{\rlap{\lower5pt\hbox{$\sim$}}}}
\def\udtw{\smash{\rlap{\lower6pt\hbox{$\approx$}}}}

\title[Limb-darkening coefficients from non-LTE model atmospheres] {Limb-darkening coefficients from line-blanketed non-LTE
  hot-star model atmospheres} \author[D. C. Reeve and I. D. Howarth ]
{D. C. Reeve and I. D. Howarth\thanks{E-mail: i.howarth@ucl.ac.uk} \\
  Dept. of Physics and Astronomy, University College London, Gower
  Street, London WC1E 6BT, UK }
\begin{document}

\date{Accepted. Received; in original form}


\maketitle

\label{firstpage}

\begin{abstract}
  We present grids of limb-darkening coefficients computed from non-LTE,
  line-blanketed {\sc tlusty} model atmospheres, covering
  effective-temperature and surface-gravity ranges of 15--55~kK and
  4.75~dex (cgs) down to the effective Eddington limit, at 2$\times$,
  1$\times$, 0.5$\times$ (LMC), 0.2$\times$ (SMC), and 0.1$\times$
  solar.  Results are given for the Bessell \emph{UBVRI$_{\rm C}$JKHL}, Sloan
  \emph{ugriz}, Str\"omgren \emph{ubvy}, WFCAM \emph{ZYJHK},
  \emph{Hipparcos, Kepler,} and \emph{Tycho} passbands, in each case
  characterized by several different limb-darkening `laws'.  We
  examine the sensitivity of limb darkening to temperature, gravity,
  metallicity, micro\-turbulent velocity, and wavelength, and make a
  comparison with LTE models.   The dependence on metallicity is very
  weak, but limb darkening is a moderately strong function of
  \logg\ in this temperature regime.
\end{abstract}

\begin{keywords}
stars: atmospheres -- radiative transfer
\end{keywords}

\section{Introduction}
An accurate description of limb darkening is essential to any investigation
involving the direct or indirect spatial resolution of a stellar surface.
Modelling of eclipsing-binary light-curves has been a major
application historically, but studies of exoplanetary transits and
microlensing events, and inter\-ferometric observations of stellar
surfaces, have similar requirements.

The advent of good-quality line-blanketed LTE model atmospheres
(particularly from Kurucz's {\sc Atlas} codes; \citealt{Kurucz79})
made it possible to generate large grids of synthetic spectra, and
associated limb-darkening coefficients, for wide ranges of
stellar parameters (e.g., \citealt{Wade85}; \citealt{Claret00};
\citealt{Howarth11a}, and references therein).  Results for non-LTE
models are fewer, for reasons of computational
complexity and expense, and available grids extend only to $\teff\lesssim$10~kK
(\citealt{Claret03, Claret12, Claret13}).  

Here we report results from line-blanketed non-LTE models for the
OB-star regime.   These may be of particular interest in the
study of extragalactic eclipsing-binary systems, where hot, luminous
stars present good opportunities for direct distance-scale
determinations, in addition to measurements of basic stellar
parameters (e.g., \citealt{harries03, hilditch05, bonanos09,
  North10}).    They should also have applicability in accurate 
photometric modelling of hot stars that
depart from spherical symmetry, through rapid rotation 
(as in the case of Be stars; e.g., \citealt{Townsend04}),
pulsation, or surface inhomogeneities
(e.g., \citealt{Rami14, Buyss15}).

\begin{figure*}
\begin{center}
\includegraphics[angle=00, scale=0.9]{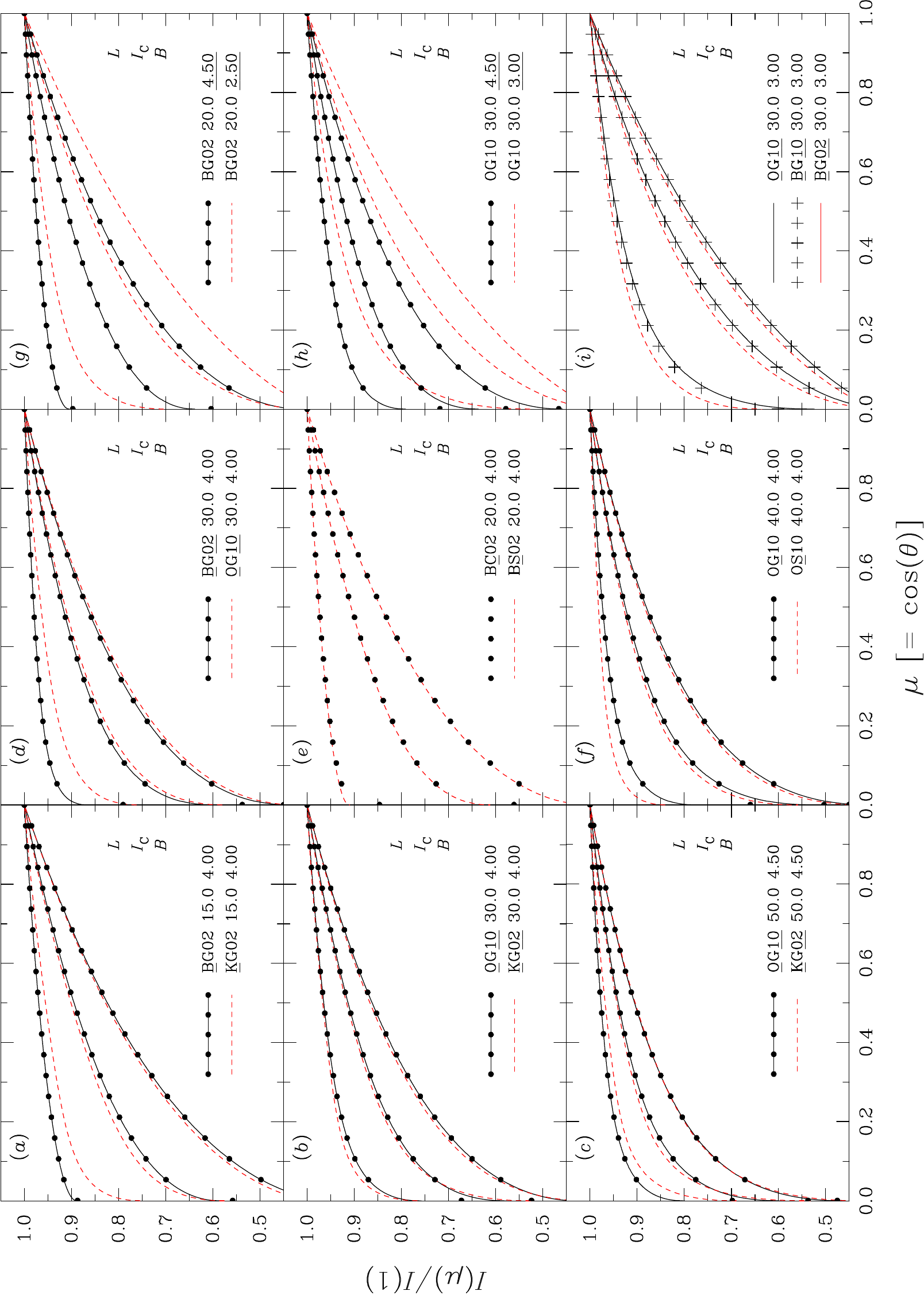} 
\caption{Representative intensity distributions to 
illustrate the sensitivity of computed
limb darkening to 
temperature (compare, e.g., panels $a$--$c$), 
gravity (panels $g$, $h$), 
microturbulence ($d$, $i$), 
abundance ($e$, $f$),
and wavelength (Johnson \emph{B, I$_{\rm C}$, \emph{and} L} bands throughout).
  Models are identified by source (B, O are non-LTE {\sc tlusty
  Bstar2006} and {\sc Ostar2002} results; K denotes {\sc Atlas}
  [`Kurucz'] LTE calculations, from \citealt{Howarth11a}),
  microturbulence (2 or 10~\kms), abundance (C, G, S for $Z/Z_\odot =
  2, 1, 0.2$), effective temperature (in~kK), and \logg\ (cgs).
  Continuous and dashed curves are parametrized (4-coefficient) characterizations, with dots
  (and `+' symbols) showing individual computed $I(\mu)$ values.  [The centre
  panel does correctly display results of two different, but almost
  indistinguishable, models.] See $\S$\ref{disco:sec} for discussion.}
\label{fig1}
\end{center}
\end{figure*}

\section{Method}
\subsection{Models}

Our calculations make use of line-blanketed {\sc tlusty} models
\citep{Hubeny95}; these assume plane-parallel, hydro\-static
atmospheres, relaxing the  approximation of local thermo\-dynamic
equilibrium.  Two extensive grids of structures are available, based
on these models: {\sc Ostar2002} (\citealt{Lanz03}; 27.5--55.0~kK)
and {\sc Bstar2006} (\citealt{Lanz07}; 15.0--30.0~kK), each
extending over a range of surface gravities from 4.75~dex~(cgs) to the
effective Eddington limit, at several metallicities.

\begin{table*}
\caption{Summary of models; sampling in gravity is at steps of 0.25~dex.
Metallicity codings are identified in
  columns 7--11 (e.g., the L grid corresponds to $Z/Z_\odot = 0.5$);  
$\xi$ is the micro\-turbulence; $N$ is
  the number of models. The high-$\xi$ models summarized in the final
  row of the Table are
  intended to correspond to B~supergiants, and do not extend to
  main-sequence gravities (cf.\ \citealt{Lanz07});  in view of the
  weak sensitivity of results to metallicity, we computed only
  solar-metallicity LDCs for these models.}
\begin{center}
\begin{tabular}{lccccccccccc}
\hline\hline
Grid & \teff & \multicolumn{2}{c}{\logg\ (min)}& (max) & $\xi$&
\multicolumn{5}{c}{$Z/Z_\odot$ coding} &$N$\\
     &  (kK) & [\teff(lo)] &\teff(hi)] & & (\kms)&\multicolumn{5}{c}{--------------------------------}\\
OStar02 & 27.5--55.0 {@} 2.5 & 3.00 & 4.00 & 4.75 &10& 2& 1& 0.5&  0.2& 0.1& 345 \\
BStar06 & 15.0--30.0 {@} 1.0 & 1.75 & 3.00 & 4.75 &2& C&G&L&S&T & 817 \\
BStar06 & 15.0--30.0 {@} 1.0 & 1.75 & 3.00 & 3.00 &10&  &G& & &  & $\phantom{3}$49 \\
\hline
\end{tabular}
\end{center}
\label{t_models}
\end{table*}

We computed emergent intensities from the atmospheric structures with
moderately dense wavelength sampling (median interval $\sim$0.4\AA) at
20 angles, approximately equally spaced over the range 0.001--1 in
$\mu \equiv \cos\theta$, where $\theta$ is the emergent angle measured
from the surface normal.  The calculations were made using Hubeny's
{\sc synspec} code,\footnote{{\tt
    http://nova.astro.umd.edu/Synspec49/synspec.html}} with the
accompanying line lists augmented with data from Kurucz's {\sc gfall}
compilation,\footnote{{\tt
    http://kurucz.harvard.edu/linelists/gfall/}} to extend coverage of
\mbox{(near-)IR} wavelengths.

Results were used to generate broad-band specific intensities
appropriate for photon-counting detectors (such as CCDs),
\begin{equation*} 
     I_i(\mu) = \frac{\int_{\lambda} I_{\lambda}(\mu)\; \phi_{i}(\lambda)\;\lambda \;\text{d}\lambda}
                     {\int_{\lambda} \phi_{i}(\lambda)\; \lambda  \;\text{d}\lambda}
\end{equation*}
where $\phi_i(\lambda)$ is the response function for passband $i$.  We
employed the response functions adopted by \citet{Howarth11a} for the
Bessell \emph{UBVRI$_{\rm C}$JKHL}, Sloan \emph{ugriz}, Str\"omgren \emph{ubvy},
WFCAM \emph{ZYJHK}, \emph{Hipparcos, Kepler,} and \emph{Tycho}
passbands.

\begin{figure}
\begin{center}
\includegraphics[angle=270, scale=0.6]{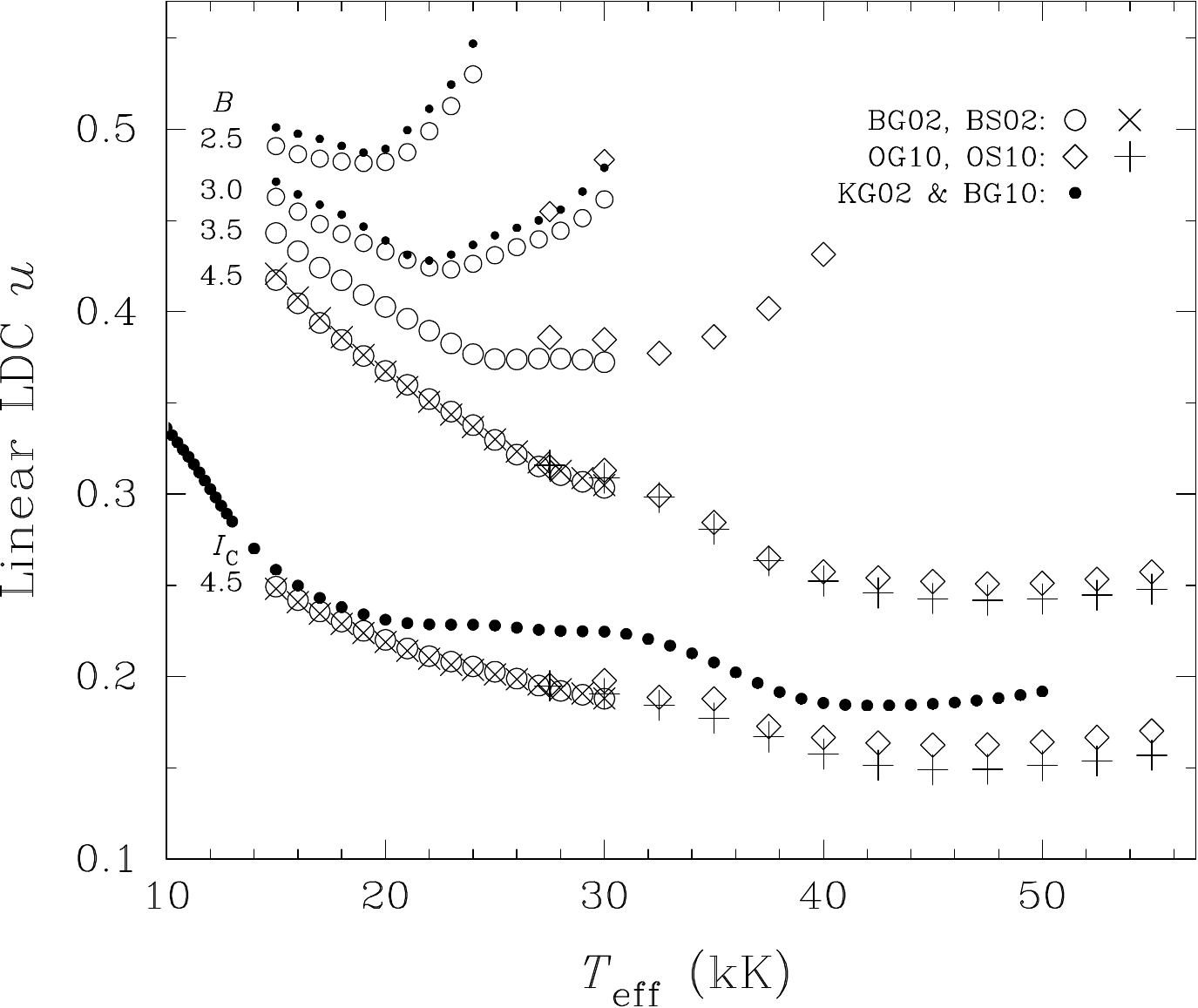} 
\caption{Linear (one-parameter) limb-darkening coefficients for
selected models, to 
  illustrate  \logg, $\xi$, and metallicity dependences.
Results are labelled by Johnson passband and \logg\  (additional results
are  available at intermediate gravities); `G' and `S' models
have $Z/Z_\odot = 1$ and 0.2, respectively;  `K' identifies LTE
(Kurucz) results from \citet{Howarth11a}.  $B$-band results 
for B-star models at $\log{g}=2.5$ and 3.0 are shown
for both
$\xi = 2$ \&
10~\kms (`normal' and `supergiant' values).}
\label{fig2}
\end{center}
\end{figure}

\subsection{Characterization}
\label{sec_ldc}

Particularly for photometric applications, limb darkening is, in
practice, conveniently characterized by ad hoc analytical `laws'
representing the specific intensities (or monochromatic radiances),
thereby reducing the data description to a small set of
limb-darkening coefficients (LDCs).

While numerous limb-darkening laws have been proposed (cf., e.g.,
\citealt{Claret00, Howarth11a}), current observations
(particularly of exoplanetary transits) are generally capable of
usefully constraining only one- or two-parameter forms. 
The analytical solution for a model atmosphere in which the source
function is linear in optical depth is
\begin{equation} 
{I_\lambda(\mu)} = {I_\lambda(1)}\left[{1 - u_\lambda(1 - \mu)}\right]
\label{eq_lin}
\end{equation}
and use of this linear limb-darkening law persists in the context of
light-curve modelling, although it often gives a rather poor
representation of model-atmosphere emergent intensities.  A quadratic law,
\begin{equation} 
{I_\lambda(\mu)} = {I_\lambda(1)}\left[{1 - u_{1,\lambda}(1 - \mu) 
- u_{2,\lambda}(1 - \mu)^2}\right] 
\label{eq_quad}
\end{equation}
offers greater flexibility, and two coefficients 
represent the practical limit for empirical determinations of LDCs
(within limits;  cf., e.g., \citealt{Howarth11b, Muller13}).
The most accurate algebraic representation of model-atmosphere results
in routine use 
is the 4-parameter law introduced by \citet{Claret00},
\begin{equation} 
{I_{\lambda}(\mu)}= 
{I_{\lambda}(1)}\left[{ 1 -
        \sum\limits_{k=1}^4 a_{k,\lambda}(1 - \mu^{k/2}) }\right],
\label{eq_cl4}
\end{equation}
which is able to fit intensity profiles across a wide parameter space
with good accuracy.

\subsection{Results}

The 1\,211 models for which we have computed emergent
intensities are summarized in Table~\ref{t_models}.  
For each model we
calculated LDCs for each `law' (eqtns.~\ref{eq_lin}--\ref{eq_cl4}) by
least-squares fits of the given functions to $I_i(\mu)$ over the range
\mbox{$0.05\le\mu\le1.00$},\footnote{We discarded $I_i(10^{-3})$, as
the  intensity can drop off rapidly at the extreme limb (cf.~Fig.~\ref{fig1}), `throwing' the fit.  This
  has no important consequences for most applications.} with and
without $I_i(1)$ as a free parameter in the fit.  We also
computed flux-conserving least-squares coefficients for the linear and
quadratic laws (following \citealt{Howarth11b}).
Results are presented on-line (q.v. Appendix~\ref{Appx1}).

\section{Discussion and Conclusions} 
\label{disco:sec}

Figure~\ref{fig1} shows the general form of the limb darkening at
representative values of temperature, gravity, abundance, and
wavelength; and illustrates the sensitivity to technical details of
the different models ({\sc Ostar2002}, {\sc Bstar2006}, and
line-blanketed LTE models from
\citealt{Howarth11a}).   These aspects are reviewed
across a  broader parameter space, but in a coarser manner,
in Fig.~\ref{fig2}, which
shows the linear limb-darkening coefficient for a range of models
(eqtn.~\ref{eq_lin}, $I(1)$ fixed; see also Fig.~1 of
\citealt{Howarth11b}).  

As well as the obvious dependences on temperature and passband
(generally, smaller limb-darkening at longer wavelengths), there is a
fairly strong dependence on gravity in the OB-star regime
(Fig.~\ref{fig2}; Fig.~\ref{fig1}, panels \textit{g, h}), although the
sensitivity to metallicity is rather weak (panels \textit{e, f}).

Differences in limb-darkening between Kurucz (LTE) and {\sc tlusty}
(non-LTE) models turn out to be reasonably small (Fig.~\ref{fig1},
panels $a$--$c$).  As expected, the {\sc Bstar2006} models at $\xi =
10$~\kms\ are in excellent agreement with their {\sc Ostar2002}
counterparts in their limited overlap range, and both show only small
differences from the $\xi = 2$~\kms\ B-star models
(Fig.~\ref{fig1}\textit{i}).

Finally, it should be borne in mind
that the LDCs presented here are all based on hydrostatic, plane-parallel
model structures.  Real O stars, and B supergiants (at
least), have substantial stellar winds, which raises the question of the
applicability of our results.  \citet{Lanz03} address this issue at
some length, and conclude that {\sc tlusty} models give a satisfactory
representation of most spectral lines in the UV--IR regime (a
conclusion that applies \textit{a fortiori} to the continuum), and
that line blanketing is the more important consideration.  We
therefore expect our broad-band LDCs to provide a reasonable
description of nature, other than when the continuum forms in
dynamical layers or where spherical extension is significant (e.g., in
Luminous Blue Variables).  These circumstances probably occur only
outside the \logg\ domain covered by the grids.

\section*{Acknowledgments}
We thank Ivan Hubeny for his support of {\sc tlusty} and
{\sc synspec}.   Our referee, Sergio Sim\'on-D{\'\i}az, 
made several useful suggestions that improved the presentation.

\bibliographystyle{mnras}
\bibliography{IDH}

\begin{thebibliography}{}
\makeatletter
\relax
\def\mn@urlcharsother{\let\do\@makeother \do\$\do\&\do\#\do\^\do\_\do\%\do\~}
\def\mn@doi{\begingroup\mn@urlcharsother \@ifnextchar [ {\mn@doi@}
  {\mn@doi@[]}}
\def\mn@doi@[#1]#2{\def\@tempa{#1}\ifx\@tempa\@empty \href
  {http://dx.doi.org/#2} {doi:#2}\else \href {http://dx.doi.org/#2} {#1}\fi
  \endgroup}
\def\mn@eprint#1#2{\mn@eprint@#1:#2::\@nil}
\def\mn@eprint@arXiv#1{\href {http://arxiv.org/abs/#1} {{\tt arXiv:#1}}}
\def\mn@eprint@dblp#1{\href {http://dblp.uni-trier.de/rec/bibtex/#1.xml}
  {dblp:#1}}
\def\mn@eprint@#1:#2:#3:#4\@nil{\def\@tempa {#1}\def\@tempb {#2}\def\@tempc
  {#3}\ifx \@tempc \@empty \let \@tempc \@tempb \let \@tempb \@tempa \fi \ifx
  \@tempb \@empty \def\@tempb {arXiv}\fi \@ifundefined
  {mn@eprint@\@tempb}{\@tempb:\@tempc}{\expandafter \expandafter \csname
  mn@eprint@\@tempb\endcsname \expandafter{\@tempc}}}

\bibitem[\protect\citeauthoryear{{Bonanos} et~al.,}{{Bonanos}
  et~al.}{2009}]{bonanos09}
{Bonanos} A.~Z.,  et~al., 2009, \mn@doi [\aj] {10.1088/0004-6256/138/4/1003},
  \href {http://adsabs.harvard.edu/abs/2009AJ....138.1003B} {138, 1003}

\bibitem[\protect\citeauthoryear{{Buysschaert} et~al.,}{{Buysschaert}
  et~al.}{2015}]{Buyss15}
{Buysschaert} B.,  et~al., 2015, \mn@doi [\mnras] {10.1093/mnras/stv1572},
  \href {http://adsabs.harvard.edu/abs/2015MNRAS.453...89B} {453, 89}

\bibitem[\protect\citeauthoryear{{Claret}}{{Claret}}{2000}]{Claret00}
{Claret} A.,  2000, \aap, \href
  {http://adsabs.harvard.edu/abs/2000A%26A...363.1081C} {363, 1081}

\bibitem[\protect\citeauthoryear{{Claret} \& {Hauschildt}}{{Claret} \&
  {Hauschildt}}{2003}]{Claret03}
{Claret} A.,  {Hauschildt} P.~H.,  2003, \mn@doi [\aap]
  {10.1051/0004-6361:20031405}, \href
  {http://adsabs.harvard.edu/abs/2003A%26A...412..241C} {412, 241}

\bibitem[\protect\citeauthoryear{{Claret}, {Hauschildt}  \& {Witte}}{{Claret}
  et~al.}{2012}]{Claret12}
{Claret} A.,  {Hauschildt} P.~H.,   {Witte} S.,  2012, \mn@doi [\aap]
  {10.1051/0004-6361/201219849}, \href
  {http://adsabs.harvard.edu/abs/2012A%26A...546A..14C} {546, A14}

\bibitem[\protect\citeauthoryear{{Claret}, {Hauschildt}  \& {Witte}}{{Claret}
  et~al.}{2013}]{Claret13}
{Claret} A.,  {Hauschildt} P.~H.,   {Witte} S.,  2013, \mn@doi [\aap]
  {10.1051/0004-6361/201220942}, \href
  {http://adsabs.harvard.edu/abs/2013A%26A...552A..16C} {552, A16}

\bibitem[\protect\citeauthoryear{{Harries}, {Hilditch}  \& {Howarth}}{{Harries}
  et~al.}{2003}]{harries03}
{Harries} T.~J.,  {Hilditch} R.~W.,   {Howarth} I.~D.,  2003, \mn@doi [\mnras]
  {10.1046/j.1365-8711.2003.06169.x}, \href
  {http://adsabs.harvard.edu/abs/2003MNRAS.339..157H} {339, 157}

\bibitem[\protect\citeauthoryear{{Hilditch}, {Howarth}  \&
  {Harries}}{{Hilditch} et~al.}{2005}]{hilditch05}
{Hilditch} R.~W.,  {Howarth} I.~D.,   {Harries} T.~J.,  2005, \mn@doi [\mnras]
  {10.1111/j.1365-2966.2005.08653.x}, \href
  {http://adsabs.harvard.edu/abs/2005MNRAS.357..304H} {357, 304}

\bibitem[\protect\citeauthoryear{{Howarth}}{{Howarth}}{2011a}]{Howarth11a}
{Howarth} I.~D.,  2011a, \mn@doi [\mnras] {10.1111/j.1365-2966.2011.18122.x},
  \href {http://adsabs.harvard.edu/abs/2011MNRAS.413.1515H} {413, 1515}

\bibitem[\protect\citeauthoryear{{Howarth}}{{Howarth}}{2011b}]{Howarth11b}
{Howarth} I.~D.,  2011b, \mn@doi [\mnras] {10.1111/j.1365-2966.2011.19568.x},
  \href {http://adsabs.harvard.edu/abs/2011MNRAS.418.1165H} {418, 1165}

\bibitem[\protect\citeauthoryear{{Hubeny} \& {Lanz}}{{Hubeny} \&
  {Lanz}}{1995}]{Hubeny95}
{Hubeny} I.,  {Lanz} T.,  1995, \mn@doi [\apj] {10.1086/175226}, \href
  {http://adsabs.harvard.edu/abs/1995ApJ...439..875H} {439, 875}

\bibitem[\protect\citeauthoryear{{Kurucz}}{{Kurucz}}{1979}]{Kurucz79}
{Kurucz} R.~L.,  1979, \mn@doi [\apjs] {10.1086/190589}, \href
  {http://adsabs.harvard.edu/abs/1979ApJS...40....1K} {40, 1}

\bibitem[\protect\citeauthoryear{{Lanz} \& {Hubeny}}{{Lanz} \&
  {Hubeny}}{2003}]{Lanz03}
{Lanz} T.,  {Hubeny} I.,  2003, \mn@doi [\apjs] {10.1086/374373}, \href
  {http://adsabs.harvard.edu/abs/2003ApJS..146..417L} {146, 417 (erratum in
  147, 225)}

\bibitem[\protect\citeauthoryear{{Lanz} \& {Hubeny}}{{Lanz} \&
  {Hubeny}}{2007}]{Lanz07}
{Lanz} T.,  {Hubeny} I.,  2007, \mn@doi [\apjs] {10.1086/511270}, \href
  {http://adsabs.harvard.edu/abs/2007ApJS..169...83L} {169, 83}

\bibitem[\protect\citeauthoryear{{M{\"u}ller}, {Huber}, {Czesla}, {Wolter}  \&
  {Schmitt}}{{M{\"u}ller} et~al.}{2013}]{Muller13}
{M{\"u}ller} H.~M.,  {Huber} K.~F.,  {Czesla} S.,  {Wolter} U.,   {Schmitt}
  J.~H.~M.~M.,  2013, \mn@doi [\aap] {10.1051/0004-6361/201322079}, \href
  {http://adsabs.harvard.edu/abs/2013A%26A...560A.112M} {560, A112}

\bibitem[\protect\citeauthoryear{{North}, {Gauderon}, {Barblan}  \&
  {Royer}}{{North} et~al.}{2010}]{North10}
{North} P.,  {Gauderon} R.,  {Barblan} F.,   {Royer} F.,  2010, \mn@doi [\aap]
  {10.1051/0004-6361/200810284}, \href
  {http://adsabs.harvard.edu/abs/2010A%26A...520A..74N} {520, A74 (corrigendum
  in 540, C1)}

\bibitem[\protect\citeauthoryear{{Ramiaramanantsoa} et~al.,}{{Ramiaramanantsoa}
  et~al.}{2014}]{Rami14}
{Ramiaramanantsoa} T.,  et~al., 2014, \mn@doi [\mnras] {10.1093/mnras/stu619},
  \href {http://adsabs.harvard.edu/abs/2014MNRAS.441..910R} {441, 910}

\bibitem[\protect\citeauthoryear{{Townsend}, {Owocki}  \& {Howarth}}{{Townsend}
  et~al.}{2004}]{Townsend04}
{Townsend} R.~H.~D.,  {Owocki} S.~P.,   {Howarth} I.~D.,  2004, \mn@doi
  [\mnras] {10.1111/j.1365-2966.2004.07627.x}, \href
  {http://adsabs.harvard.edu/abs/2004MNRAS.350..189T} {350, 189}

\bibitem[\protect\citeauthoryear{{Wade} \& {Rucinski}}{{Wade} \&
  {Rucinski}}{1985}]{Wade85}
{Wade} R.~A.,  {Rucinski} S.~M.,  1985, \aaps, \href
  {http://adsabs.harvard.edu/abs/1985A%26AS...60..471W} {60, 471}

\makeatother
\end{thebibliography}

\onecolumn
\appendix

\section[]{On-line material}
\label{Appx1}

Results are given in two summary files (one for linear and
quadratic forms  and one for 4-coefficient fits),  and 
for individual models, identifiable by file name;
e.g., the file {\tt BG20000g425v02.LDC} is from the Bstar06, G-abundance
(`Galactic', $Z/Z_\odot = 1$) grid, for $\teff = 20$~kK,
$\logg = 4.25$, micro\-turbulence $\xi = 2$~\kms.
Each individual model file contains a series of blocks giving results
for different passbands;  for example, \texttt{BG20000g425v02.LDC}  starts
\scriptsize
\begin{verbatim}
#-------------------------------------------------------------------------------------------------------------------------
# Bessell-UX          3598.5  5.99074E+08  2.15116E+08
  linear - 1    +1.00000E+00 +3.82029E-01                                            +2.463E-02 -6.659E-02  0.05    +1.557
  linear - 2    +1.02778E+00 +4.24870E-01                                            +1.995E-02 -5.383E-02  0.05    +0.034
  lin - FCLS    +1.02822E+00 +4.13609E-01                                            +1.996E-02 -5.387E-02  0.05     0.000
  quad - 2      +1.00000E+00 +2.16789E-01 +2.26667E-01                               +7.099E-03 -1.995E-02  0.05    -0.395
  quad - 3      +9.91711E-01 +1.82744E-01 +2.55834E-01                               +6.406E-03 +1.776E-02  0.05    -0.192
  quad - FCLS   +9.89536E-01 +1.84220E-01 +2.56568E-01                               +6.584E-03 -1.776E-02  0.05     0.000
  4-coeff       +1.00000E+00 +5.16742E-01 +5.48940E-02 -1.82686E-02 -2.91968E-03     +1.815E-05 +5.219E-05  0.11    -0.134
  1+4-coeff     +9.99981E-01 +5.17614E-01 +5.22650E-02 -1.50032E-02 -4.34753E-03     +1.729E-05 +4.932E-05  0.11    -0.133
#-------------------------------------------------------------------------------------------------------------------------
# Bessell-BX          4376.9  4.24681E+08  1.51335E+08
  linear - 1    +1.00000E+00 +3.73773E-01                                            +3.170E-02 -8.671E-02  0.05    +1.997
  linear - 2    +1.03522E+00 +4.28079E-01                                            +2.587E-02 -7.053E-02  0.05    +0.082
  ...
\end{verbatim}
\normalsize
The first line lists 
\begin{itemize}
\item[(i)] the passband identifier (here Bessell's characterizations of
  the Johnson \emph{U} and \emph{B} passbands used in computing
  $U-B$ colours); 
\item[(ii)] the effective wavelength (in \AA);
\item[(iii)] the broad-band physical flux $F_\lambda$ ($=4\pi H_\lambda$, 
in erg cm$^{-2}$ s$^{-1}$ \AA$^{-1}$); 
\item[(iv)] the broad-band surface-normal intensity $I_\lambda(1)$
(in erg cm$^{-2}$ s$^{-1}$ \AA$^{-1}$ sr$^{-1}$)
\end{itemize}
(multiply values by $10^{-2}$ to obtain results in units of
J m$^{-2}$ s$^{-1}$ nm$^{-1}$).
Subsequent lines list coefficients for the limb-darkening laws
discussed in Section~\ref{sec_ldc};  `{\tt FCLS}' indicates values
obtained by flux-conserving least squares.
The first value is
$\hat{I}_\lambda(1)/I_\lambda(1)$, the ratio of the fitted 
surface-normal intensity to the model-atmosphere value, followed by
the set of limb-darkening coefficients.

The last four columns are indicators of the fit quality, being the
r.m.s. and maximum O$-$C values of $I_\lambda(\mu)/I_\lambda(1)$
(where `O' and `C' refer to model-atmosphere and analytical-fit
values); the value of $\mu$ for which the absolute value of the
mismatch is greatest;  and the percentage error in the flux
obtained by integrating the analytical represention of the intensity
to obtain
the first moment of the radiation field (Eddington flux, $H_\lambda =
F_\lambda/4\pi$).

\smallskip
The summary 
files extract and compile the LDC information in a format intended to
be self-explanatory, and amenable to processing by simple
text-manipulation tools (e.g., {\sc{UNIX}} {\tt grep}, {\tt awk}, and {\tt cut}
commands).  The 
\texttt{Summary1.LDC} 
file of linear and quadratic coefficients, for
example, begins \scriptsize
\begin{verbatim}
# Source file    filter         I(1)     ------- linear 1 --------   +++++++ linear 2 ++++++++   ------- lin FCLS --------   +++++++++++++++ quad 2 +++++++++++
BC15000g175v02 Bessell-UX   1.36727E+08  +1.00000E+00 +5.46138E-01   +1.02181E+00 +5.79776E-01   +1.02255E+00 +5.67667E-01   +1.00000E+00 +4.20039E-01 +1.72...
BC15000g175v02 Bessell-BX   8.86990E+07  +1.00000E+00 +5.58679E-01   +1.02552E+00 +5.98034E-01   +1.02663E+00 +5.83534E-01   +1.00000E+00 +4.14861E-01 +1.97...
BC15000g175v02 Bessell-B    8.98503E+07  +1.00000E+00 +5.59268E-01   +1.02541E+00 +5.98454E-01   +1.02651E+00 +5.84003E-01   +1.00000E+00 +4.16113E-01 +1.96...
BC15000g175v02 Bessell-V    4.60602E+07  +1.00000E+00 +5.21743E-01   +1.03041E+00 +5.68644E-01   +1.03177E+00 +5.52366E-01   +1.00000E+00 +3.47071E-01 +2.39...
BC15000g175v02 Bessell-R    2.61041E+07  +1.00000E+00 +4.86751E-01   +1.03291E+00 +5.37504E-01   +1.03432E+00 +5.20945E-01   +1.00000E+00 +2.94742E-01 +2.63...
BC15000g175v02 Bessell-I    1.29020E+07  +1.00000E+00 +4.38498E-01   +1.03460E+00 +4.91859E-01   +1.03593E+00 +4.76000E-01   +1.00000E+00 +2.33205E-01 +2.81...
...
\end{verbatim}
\normalsize
The \texttt{Summary2.LDC} file has corresponding results for the
coefficients in eqtn.~\ref{eq_cl4}.

\bigskip\noindent
All files are intended to be available through CDS, and are presently
available at \url{ftp://ftp.star.ucl.ac.uk/idh/LDC/}.

\label{lastpage}

\end{document}